\documentclass[twocolumn,showpacs,prl]{revtex4}
\usepackage{dcolumn}
\usepackage{bm}

\begin{document}

\title{Ultra Low Momentum Neutron Catalyzed  \\ 
Nuclear Reactions on Metallic Hydride Surfaces}

\author{A. Widom}
\affiliation{Physics Department, Northeastern University, 110 Forsyth Street,
Boston MA 02115}
\author{L. Larsen}
\affiliation{Lattice Energy LLC, 175 North Harbor Drive, Chicago IL 60601}

\begin{abstract}
Ultra low momentum neutron catalyzed nuclear reactions in metallic hydride 
system surfaces are discussed. Weak interaction catalysis initially occurs 
when neutrons (along with neutrinos) are produced from the protons which capture 
``heavy'' electrons. Surface electron masses are shifted upwards by localized condensed 
matter electromagnetic fields. Condensed matter quantum electrodynamic processes 
may also shift the densities of final states allowing an appreciable production 
of extremely low momentum neutrons which are thereby efficiently absorbed by 
nearby nuclei. No Coulomb barriers exist for the weak interaction neutron 
production or other resulting catalytic processes. 
\end{abstract}

\pacs{24.60.-k, 23.20.Nx}
\maketitle

\section{Introduction \label{Intro}}

It is very well known that a proton \begin{math} p^+ \end{math} can capture 
a charged lepton \begin{math} l^-  \end{math} and produce 
a neutron and a neutrino from the resulting process\cite{Marshak:1969} 
\begin{equation}
l^- +p^+ \to n+\nu_l.
\label{lp}
\end{equation}
A common form of nuclear transmutation in condensed matter is understood in 
terms of Eq.(\ref{lp}). An electron \begin{math} e^- \end{math}
which wanders into a nucleus with \begin{math} Z \end{math} protons and 
\begin{math} N=A-Z \end{math} neutrons can be captured producing an electron 
neutrino \begin{math} \nu_e \end{math} and leaving behind a nucleus with 
\begin{math} Z-1 \end{math} protons and \begin{math} N+1=A-(Z-1) \end{math} 
neutrons. The electron capture process in a condensed matter nucleus may be 
described by the nuclear transmutation reaction\cite{Yukawa:1935,Moller:1937}
\begin{equation}
e^- + (A,Z) \to (A,Z-1) +\nu_e.
\label{ep}
\end{equation}
Note the absence of a Coulomb barrier to such a weak interaction nuclear process. 
In fact, a strong Coulomb {\em attraction} which can exist between an electron and 
a nucleus  {\em helps} the nuclear transmutation Eq.(\ref{ep}) proceed.
While the process Eq.(\ref{lp}) is experimentally known to occur when muons are 
mixed into Hydrogen systems\cite{Rothberg:1963,Quaranta:1969,Ando:2000}, i.e. 
\begin{math} \mu^- +p^+ \to n+\nu_\mu  \end{math}, it is regarded as difficult 
for nature to play the same trick with electrons and protons at virtual rest. 
For Eq.(\ref{lp}) to spontaneously occur it is required that the lepton mass obey 
a threshold condition,
\begin{equation}
M_l c^2 > M_nc^2 -M_p c^2\approx 1.293\ MeV\approx 2.531\ M_e c^2, 
\label{threshold}
\end{equation} 
which holds true by a large margin for the muon but is certainly not true 
for the vacuum mass of the electron. On the other hand, the electron mass in 
condensed matter can be modified by local electromagnetic field fluctuations. 
To see what is involved, one may employ a quasi-classical argument 
wherein the electron four momentum 
\begin{math} p_\mu =\partial_\mu S \end{math} in an electromagnetic field 
\begin{math} F_{\mu \nu}=\partial_\mu A_\nu -\partial_\nu A_\mu  \end{math}
obeys the Hamilton-Jacobi equation\cite{Landau:1975} 
\begin{equation}
-(p_\mu -\frac{e}{c}A_\mu )(p^\mu -\frac{e}{c}A^\mu )=M_e^2c^2 . 
\label{QuasiClassical}
\end{equation}
If the field fluctuations average to zero 
\begin{math} \overline{A_\mu}=0  \end{math}, 
then the remaining mean square fluctuations can on average add mass to the  
electron \begin{math} M_e \to \tilde{M}_e \end{math} according to a previously 
established rule\cite{Landau:1975,Lifshitz:1997}   
\begin{equation}
-\tilde{p}_\mu \tilde{p}^\mu =
\tilde{M}_e^2c^2=M_e^2c^2+\left(\frac{e}{c}\right)^2\overline{A^\mu A_\mu }\ .
\label{MassRenormalize}
\end{equation}
For example laser light fields can ``dress'' an electron in a 
non-perturbation theoretical fashion with an additional mass as in 
Eq.(\ref{MassRenormalize}). Such mass modifications must be applied 
to electrons and positrons when pairs can in principle be blasted out 
of the vacuum\cite{Ringwald:2001,Popov:2002}
employing colliding laser beams. The mass growth in the theory appears in a 
classic treatise on quantum electrodynamics\cite{Lifshitz:1997}. 
The theory in terms of condensed matter photon propagators is discussed below.

The mass modified hydrogen atom can decay into a neutron and a neutrino 
if the mass growth obeys a threshold condition given by
\begin{eqnarray}
\beta &\equiv & \frac{\tilde{M}_e}{M_e}
=\left[1+\left(\frac{e}{M_e c^2}\right)^2\overline{A^\mu A_\mu }\right]^{1/2}
\nonumber \\
\beta &>& 2.531 \ \ {\rm (neutron\ production)}.
\label{collapse}
\end{eqnarray}
The sources of the electron mass renormalization via electromagnetic field fluctuations 
on metallic hydride surfaces and the resulting neutron production are the main 
subject matters of this work. The surface states of metallic hydrides are of 
central importance: (i)  Collective surface plasma\cite{Stern:1960} modes are involved 
in the condensed matter weak interaction density of final states. The radiation frequencies 
of such modes range from the infrared to the soft X-ray spectra. (ii) The 
breakdown\cite{White:2005} of the conventional Born-Oppenheimer 
approximation for the surface hydrogen atoms contributes 
to the large magnitude of electromagnetic fluctuations. Some comments regarding 
nuclear transmutation reactions which result from ultra low momentum neutron production 
will conclude our discussion of neutron catalyzed reactions. 

\section{Electromagnetic Field Fluctuations \label{PhotonField}}

The rigorous definition of electron mass growth due to the metallic hydride
electromagnetic fields depends on the non-local ``self'' mass 
\begin{math} {\cal M}  \end{math} in  
the electron Green's function\cite{Schwinger:1951} \begin{math} G \end{math}, i.e. 
\begin{eqnarray}
-i\gamma^\mu \partial_\mu G(x,y) &+&
\frac{c}{\hbar }\int {\cal M} (x,z)G(z,y)d^4z=\delta (x-y),
\nonumber \\ 
{\cal M} (x,y)&=&M_e\delta (x-y)+\frac{\hbar}{c}\Sigma (x,y),
\label{SelfEnergy}
\end{eqnarray}
wherein the non-local mass shift operator \begin{math} \Sigma  \end{math} 
depends on the difference between the photon propagator 
\begin{equation}
D_{\mu \nu}(x,y)=\frac{i}{\hbar c}\left<A_\mu (x)A_\mu (y)\right>_+
\label{propagator}
\end{equation}
in the presence of condensed matter and the photon propagator 
\begin{math} D^{(0)}_{\mu \nu}(x,y)  \end{math} in the vacuum. In 
Eq.(\ref{propagator}), ``+'' denotes time ordering. The source of 
the differences in the photon propagators  
\begin{eqnarray} 
&\ & \ \ \ \ \ \
\ \ \ \ \ \ D_{\mu \nu}(x,y)-D^{(0)}_{\mu \nu}(x,y)=
\nonumber \\
&\ & \int \int  D^{(0)}_{\mu \sigma }(x,x^\prime )
{\cal P}^{\sigma \lambda }(x^\prime ,y^\prime )
D^{(0)}_{\lambda \nu}(y^\prime ,y)d^4 x^\prime d^4 y^\prime 
\label{polarization}
\end{eqnarray}
is the polarization response function 
\begin{math} {\cal P}^{\sigma \lambda }(x,y)  \end{math}
arising from condensed matter currents 
\begin{equation}
{\cal P}^{\mu \nu }(x,y)=\frac{i}{\hbar c^3}
\left<J^\mu (x)J^\nu (y)\right>_+ .
\label{currents}
\end{equation}
The gauge invariant currents in Eq.(\ref{currents}) give rise to the 
electromagnetic fluctuations which only at first sight appear not 
to be gauge invariant. The average of the field fluctuations appearing 
in Eq.(\ref{collapse}) is in reality what is obtained after subtracting 
the vacuum field fluctuations which partially induce the physical 
vacuum electron mass; i.e. 
\begin{eqnarray}
\overline{A^\mu (x)A_\mu (x)} &=& \left<A^\mu (x)A_\mu (x)\right>
-\left<A^\mu (x)A_\mu (x)\right>_{vac}\ ,
\nonumber \\  
\frac{i}{\hbar c}\overline{A^\mu (x)A_\mu (x)} &=& 
D^{\mu }_{\ \mu }(x,x)-
D^{(0) \mu }_{\ \ \ \ \ \mu }(x,x) . 
\label{fluctuation1}
\end{eqnarray}
In terms of the spectral function 
\begin{math} S({\bf r},\omega ) \end{math} defined by the electric field 
anti-commutator 
\begin{equation}
2\int_{-\infty}^\infty S_{\bf EE}({\bf r},\omega )\cos(\omega t)d\omega 
=\overline{\{{\bf E}({\bf r},t);{\bf E}({\bf r},0)\}},
\label{fluctuation2}
\end{equation}
the local electronic mass enhancement factor Eq.(\ref{collapse}) is given by 
\begin{equation}
\beta ({\bf r})
=\left[1+\left(\frac{e}{M_e c}\right)^2
\int_{-\infty}^\infty S_{\bf EE}({\bf r},\omega )\frac{d\omega }{\omega^2}\right]^{1/2}.
\label{fluctuation3}
\end{equation}
The frequency scale \begin{math} \tilde{\Omega } \end{math} of the electric field 
oscillations may be defined via  
\begin{equation}
\frac{1}{\tilde{\Omega}^2}\overline{|{\bf E}({\bf r})|^2}\equiv
\int_{-\infty}^\infty S_{\bf EE}({\bf r},\omega )\frac{d\omega }{\omega^2}\ ,
\label{frequency4}
\end{equation}
so that 
\begin{equation}
\beta ({\bf r})
= \sqrt{1+\frac{\overline{|{\bf E}({\bf r})|^2}}{{\cal E}^2}}
\ \ {\rm wherein} 
\ \ {\cal E} = \left|\frac{M_ec\tilde{\Omega}}{e}\right|
\label{frequency5}
\end{equation}
which is an obviously gauge invariant result. When an electron wanders into a proton 
to produce a neutron and a neutrino, the electric fields forcing oscillations of the 
electrons are largely due to the protons themselves. Considerable experimental information 
about the proton oscillations in metallic hydride systems is available 
from neutron beams scattering off protons.

\section{Proton Oscillations \label{ProOsc}}

A neutron scattering from \begin{math} N \end{math} protons in metallic hydride systems 
probes the quantum oscillations of protons as described by the correlation 
function\cite{Squires:1996}  
\begin{equation}
G({\bf Q},\omega )=\frac{1}{N}\sum_{k=1}^N 
\int_{-\infty}^\infty \left<e^{-i{\bf Q\cdot R}_k(t)}
e^{i{\bf Q\cdot R}_k(0)}\right>\frac{dt}{2\pi }\ .
\label{ProOsc1}
\end{equation}
Here, \begin{math} {\bf R}_k(t) \end{math} is the position of the 
\begin{math} k^{th} \end{math} proton at time \begin{math} t \end{math}.  
The differential extinction coefficient for a neutron to scatter from the 
metallic hydride with momentum transfer 
\begin{math} \hbar {\bf Q}={\bf p}_i - {\bf p}_f \end{math} and energy transfer 
$\hbar \omega =\epsilon_i -\epsilon_f$ is given by 
\begin{equation}
\frac{d^2 h}{d\Omega_f d\epsilon_f}\approx 
\frac{\bar{\rho }}{\hbar } \left[\frac{d\sigma }{d\Omega_f }\right]G({\bf Q},\omega ),
\label{ProOsc2}
\end{equation}
wherein \begin{math} \bar{\rho } \end{math} is the mean number of protons per 
unit volume and \begin{math} d\sigma \end{math} is the elastic differential cross 
section for a neutron to scatter off a single proton into a final solid angle 
\begin{math} d\Omega_f \end{math}. 

While the weak interaction neutron production  
may occur for a number of metallic hydrides, palladium hydrides are particularly well 
studied. For a highly loaded hydride, there will be a full proton layer on the hydride 
surface. The frequency scale \begin{math} \tilde{\Omega} \end{math} of oscillating 
surface protons may be computed on the basis of neutron 
scattering data\cite{Hauer:2004,Kemali:2000}. The electric field scale 
in Eq.(\ref{frequency5}) may be estimated by 
\begin{equation}
{\cal E}\approx 1.4\times 10^{11}\ {\rm volts/meter} 
\ \ ({\rm Hydrogen\ Monolayer}).
\label{ProOsc3}
\end{equation}
The magnitude of the electric field impressed on the electronic system due to the collective 
proton layer oscillations on the surface of the palladium may be estimated by 
\begin{equation}
\sqrt{\overline{|{\bf E}|^2}}\approx \frac{4|e|\sqrt{\overline{|{\bf u}|^2}}}{3a^3}  
\ \ ({\rm Hydrogen\ Monolayer}).
\label{ProOsc4}
\end{equation}
where \begin{math} {\bf u} \end{math} is the displacement of the collective proton 
oscillations and the Bohr radius is given by 
\begin{equation}
a=\frac{\hbar^2 }{e^2M_e}\approx 0.5292\times 10^{-8}\ {\rm cm}.
\label{ProOsc5}
\end{equation}
Thus  
\begin{equation}
\sqrt{\overline{|{\bf E}|^2}}\approx  6.86\times 10^{11} ({\rm volts/meter})
\sqrt{\frac{\overline{|{\bf u}|^2}}{a^2}}\ .  
\label{ProOsc6}
\end{equation}
One may again appeal to neutron scattering from protons in palladium for the 
room temperature estimate 
\begin{equation}
\sqrt{\frac{\overline{|{\bf u}|^2}}{a^2}}
\approx 4.2   
\ \ ({\rm Hydrogen\ Monolayer}).
\label{ProOsc7}
\end{equation}
From Eqs.(\ref{frequency5}), (\ref{ProOsc3}), (\ref{ProOsc6}) and (\ref{ProOsc7}) 
follows the electron mass enhancement
\begin{equation}
\beta \approx 20.6 \ \ ({\rm Palladium\ Hydride\ Surface}).
\label{ProOsc8}
\end{equation}
The threshold criteria derived from Eq.(\ref{collapse}) is 
satisfied. On palladium, surface protons can capture a heavy 
electron producing an ultra low momentum neutron plus a neutrino; i.e. 
\begin{equation} 
(e^-p^+)\equiv H\to n+\nu_e.
\label{collapse1}
\end{equation}

Several comments are worthy of note: (i) The collective proton motions  
for a completed hydrogen monolayer on the Palladium surface require a loose coupling 
between electronic surface plasma modes and the proton oscillation modes. The 
often assumed Born Oppenheimer approximation is thereby violated. This is in fact 
the usual situation for surface electronic states as has been recently discussed. 
It is not possible for electrons to follow the nuclear vibrations on surfaces very 
well since the surface geometry precludes the usual very short Coulomb screening lengths. 
(ii) The above arguments can be extended to heavy hydrogen 
\begin{math} (e^-p^+n)\equiv (e^-d^+)\equiv D\end{math} 
wherein the neutron producing heavy electron capture has the threshold electron mass 
enhancement 
\begin{equation}
\frac{\tilde{M}_e^\prime }{M_e}=\beta^\prime (D\to n+n+\nu_e) > 6.88.  
\label{collapse2}
\end{equation}
Eq.(\ref{collapse2}) also holds true. The value of \begin{math} \beta  \end{math} 
in Eq.(\ref{ProOsc8}) is similar in magnitude for both the proton and  
the deuterium oscillation cases at hand. Since each deuterium electron capture  
yields two ultra low momentum neutrons, the nuclear catalytic reactions are somewhat more 
efficient for the case of deuterium. (iii) However, one seeks to have either nearly pure 
proton or nearly pure deuterium systems since only the isotopically pure systems will easily 
support the required coherent collective oscillations. (iv) An enforced chemical 
potential difference or pressure difference across a palladium surface 
will pack the surface layer to a single compact layer allowing for the 
required coherent electric field producing motions. 
(v) The proton electric field producing oscillations can be amplified by inducing 
an enhancement in the weakly coupled electronic surface plasma modes. Thus, 
appropriate frequencies of laser light incident on a palladium surface launching 
surface plasma waves can enhance the production of catalytic neutrons. (vi) The captured electron 
is removed from the collective surface plasma oscillation creating a large 
density of final states for the weak interactions. 
Most of the heat of reaction is to be found in these surface electronic modes. 
(vii) The neutrons themselves are produced at very low momenta, or equivalently, with 
very long wavelengths. Such neutrons exhibit very large absorption cross sections 
which are inversely proportional to neutron velocity. 
Very few of such neutrons will escape the immediate vicinity. These will rarely  
be experimentally detected. In this regard, ultra low momentum neutrons 
may produce ``neutron rich'' nuclei in substantial quantities. These neutrons 
can yield interesting reaction sequences\cite{Iwamura:2002}. Other examples are 
discussed below in the concluding section. 

\section{Low Energy Nuclear Reactions\label{lenr}}

The production of ultra low momentum neutrons can induce chains of nuclear reactions 
in neighboring condensed matter\cite{Lide:2000,Firestone:1999}. For example, let us 
suppose an initial concentration of lithium very near a suitable metallic hydride 
surface employed to impose a substantial chemical potential difference across the 
hydride surface. In that case, the existence of weak interaction 
produced surface neutrons allow for the following chain of reactions 
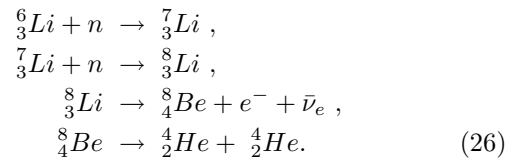
\begin{eqnarray}
 ^6 _3 Li +n &\to & ^7 _3 Li\ ,
\nonumber \\
 ^7 _3 Li +n &\to & ^8 _3 Li\ ,
\nonumber \\   
^8 _3 Li &\to & ^8 _4 Be +e^-+\bar{\nu}_e\ ,
\nonumber \\
^8 _4 Be &\to & ^4 _2 He +\ ^4 _2 He .
\label{lenr1}
\end{eqnarray}
The chain Eq.(\ref{lenr1}) yields a quite large heat \begin{math} Q \end{math} of the 
net nuclear reaction 
\begin{equation}
Q\big\{\ ^6 _3 Li +2n \to 2\ ^4 _2 He+e^-+\bar{\nu}_e\big\}\approx 26.9\ MeV. 
 \label{lenr2}
\end{equation}
Having produced \begin{math} ^4He  \end{math} products, further neutrons may be employed 
to build heavy helium ``halo nuclei'' yielding 
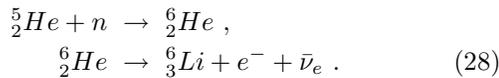
\begin{eqnarray}
 ^4 _2 He +n &\to & ^5 _2 He\ ,
\nonumber \\
 ^5 _2 He +n &\to & ^6 _2 He\ ,
\nonumber \\   
^6 _2 He &\to & ^6 _3 Li + e^- +\bar{\nu}_e\ .
\label{lenr3}
\end{eqnarray}
The chain Eq.(\ref{lenr3}) yields a moderate  
heat of the net \begin{math}  ^6 _3 Li  \end{math} producing reaction
\begin{equation}
Q\big\{\ ^4 _2 He +2n \to \ ^6 _3 Li+e^-+\bar{\nu}_e\big\}\approx 2.95\ MeV. 
 \label{lenr4}
\end{equation}
The reaction Eqs.(\ref{lenr1}) and (\ref{lenr3}) taken together 
form a nuclear reaction cycle. Other possibilities include the direct 
lithium reaction 
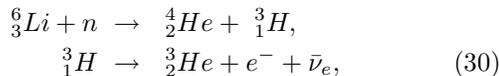
\begin{eqnarray}
^6 _3 Li +n &\to &\ ^4 _2 He+\ ^3 _1 H, 
\nonumber \\ 
^3 _1 H &\to & \ ^3 _2 He +e^- +\bar{\nu}_e,
\label{5a}
\end{eqnarray}
with the heat of net reaction 
\begin{equation}
Q \big\{\ ^6 _3 Li +n \to\  ^4 _2 He+\ ^3 _2 He +e^- +\bar{\nu}_e \big\} 
\approx  4.29\ MeV.
\label{lenr5b}
\end{equation}
This reaction yields both \begin{math} ^4 He \end{math} and 
\begin{math} ^3 He \end{math} products. All of the above reactions 
depend on the original production of neutrons. 
Of the above possible reactions, the lithium beta decay 
\begin{math} 
Q\{\ ^8 _3 Li \to \ ^8 _4 Be +e^-+\bar{\nu}_e\} \approx 16.003\ MeV  
\end{math}
in Eq.(\ref{lenr1}) yields the greatest of the above heats of nuclear fuel burning.

In summary, weak interactions can produce neutrons and neutrinos via the capture 
by protons of heavy electrons. The collective motions of the surface metallic 
hydride protons produce the oscillating electric fields which renormalize the electron 
self energy adding significantly to the effective mass. There is no Coulomb barrier 
obstruction to the resulting neutron catalyzed nuclear reactions. The final 
products \begin{math} ^A _Z X  \end{math} in some reaction chains may have 
fairly high \begin{math} A  \end{math}. The above examples show that final products 
such as \begin{math} ^4 _2 He  \end{math} do not necessarily constitute 
evidence for the direct fusion \begin{math} D+D \to \ ^4 _2 He  \end{math}. 
Direct fusion requires 
tunneling through a high Coulomb barrier. By contrast, there are no such barriers to weak 
interactions and ultra low momentum neutron catalysis. Final products such as 
\begin{math} ^4 _2 He  \end{math} and/or \begin{math} ^3 _2 He  \end{math} 
and/or \begin{math} ^3 _1 H  \end{math} may be detected.

\end{document}